\documentclass[%
 aip,
cp,  % Conference Proceedings
 amsmath,amssymb,%nobibnotes,
 reprint,%
]{revtex4-2}

\usepackage{graphicx}% Include figure files
\usepackage{dcolumn}% Align table columns on decimal point
\usepackage{bm}% bold math
\usepackage[utf8]{inputenc}
\usepackage[T1]{fontenc}
\usepackage{mathptmx}

\usepackage{booktabs} %rend les tableaux plus beaux avec /toprule, /midrule et /bottomrule

\begin{document}

\title{Acceleration of the Solar Wind by Ambipolar Electric Field}% Force line breaks with \\

\author{Viviane Pierrard} % Write as First name Surname
 \email[Corresponding author: ]{viviane.pierrard@aeronomie.be}
\author{Maximilien Péters de Bonhome}%
 \email{maximilien.peters@student.uclouvain.be}
\affiliation{
 Solar-Terrestrial Center of Excellence and Space Physics, Royal Belgian Institute for Space Aeronomy, B-1180 Brussels, Belgium
.% Force line breaks with \\ if necessary
}

%\author{Another's Name}
% \email{third.author@anotherinstitution.edu}
%\affiliation{%
%Second institution and/or address% Force line breaks with \\ if necessary
%}%
\affiliation{Center for Space Radiations, ELIC, Université Catholique de Louvain, B-1348 Louvain-La-Neuve, Belgium}

\date{\today} % It is always \today, today, but any date may be explicitly specified
              % Not printed for conference proceedings

\begin{abstract}
Kinetic exospheric models revealed that the solar wind is accelerated by an ambipolar electric field up to supersonic velocities. The presence of suprathermal Strahl electrons at the exobase can further increase the velocity to higher values, leading to profiles comparable to the observations in the fast and slow wind at all radial distances. Such suprathermal electrons are observed at large distances and recently at low distances as well. 
  Those suprathermal electrons were introduced into the kinetic exospheric model using Kappa distributions. Here, the importance of the exobase's altitude is also underlined for its ability to maintain the electric potential to a higher level for slower winds, conversely to what is induced through the effect of a lower kappa index only.  In fact, the exobase  is located at lower altitude in the coronal holes where the density is smaller than in the other regions of the corona, allowing the wind originating from the holes to be accelerated from lower distances to higher velocities.

The new observations of Parker Solar Probe (PSP) and Solar Orbiter (SolO) from launch to mid-2023  are here used to determine the characteristics of the plasma in the corona so that the model fits best to the averaged observed profiles for the slow and fast winds. The observations at low radial distances show suprathermal electrons already well present in the Strahl in the antisunward direction and a deficit in the sunward direction, confirming the exospheric feature of almost no incoming particles. 

\end{abstract}

\maketitle

\section{\label{sec:level1}Introduction}
%:\protect\\ The line break was forced \lowercase{via} \textbackslash\textbackslash}

The physical mechanisms responsible for the heating of the corona and the acceleration of the solar wind remain a hot topic of research (\cite{Meyer-Vernet-2007P, Rouillard-2021P} for reviews). Exospheric models provide a very simplified first approximation: considering only the effects of the external force, they show that the electric potential can accelerate the wind to supersonic velocities, even considering simple Maxwellian distributions for the particles at the exobase \cite{Pierrard-2012cP}. Moreover, the possible presence of suprathermal particles in the corona has important effects on the plasma temperature increase, and an enhanced population of energetic electrons accelerates the solar wind to larger bulk velocities, especially in case of a low exobase \cite{Pierrard-Lemaire-1996P, Lamy-etal-2003, Pierrard-Lazar-2010P}. This gives a natural explanation for the fast wind originating from coronal holes, where the density is lower than in the other coronal regions. Differential heating and acceleration of minor ions can also be predicted using the exospheric approach, in agreement with ion observations in the solar wind \cite{Pierrard-Lamy-2003P}. Exospheric models predict too high temperature anisotropies ($T_\parallel/T_\perp$) in comparison to  observations, which could be corrected by including effects of Coulomb collisions \cite{Pierrard-etal-2001bP}. Waves like whistlers  can also transform the velocity distributions and lead to more realistic heat fluxes \cite{Pierrard-etal-2011P, Colo23}.
Plasma turbulence and instabilities modify the characteristics of the observed distributions, and especially their temperature anisotropies and heat fluxes \cite{zhao22}. 

In his groundbreaking work on the first predictive thermally-driven solar wind model, Parker \cite{Park58} utilized magnetohydrodynamic (MHD) approach  to successfully predict high velocities even before any measurements were possible in space. The MHD approach was later complemented by the kinetic exospheric approach showing the importance of the ambipolar electric field to accelerate the wind. Other solar wind acceleration processes have been proposed, involving magnetic reconnection and waves. Some of these effects can be included in the model via the boundary conditions and diffusion coefficients in the Fokker-Planck equation \cite{Pierrard-etal-2001bP, Pierrard-etal-2016P}. 

The NASA mission Parker Solar Probe (PSP)  makes unprecedented measurements at very low distances from the Sun, allowing to discover new features such as broadband electrostatic waves  in the near-Sun solar wind \cite{Zhao22w}. It also provides the unique opportunity to study the radial evolution of the solar wind in the inner heliosphere by comparing observations with predictions from different models. A deficit in the sunward direction is observed in the PSP electron distributions at low radial distances, confirming the exospheric feature \cite{Bercic2021wh}. The deficit occurs in 60$\%-80\%$ of electron observations within 0.2 AU, and even more
frequently in plasma with low collisional age, and a more anisotropic electron core population. The
cutoff energy varies linearly with the local electron core temperature, confirming a direct relationship to the
 electric potential \cite{Halekas-etal-2021}. PSP also provides information on the presence of suprathermal electrons at very low distance \cite{Maksimovicbook-2021} and the formation of the strahl and halo \cite{Pierrard-etal-2001P, VP23}. 

In the present work, we constrain  the exospheric Kappa model presented in Section 2 to provide the best fits with the profiles of the averaged moments observed by PSP, Solar Orbiter and OMNI shown in Section 3. The correlation between proton temperatures and velocity will be discussed in section 4 in the framework of the possibility it gives to discriminate slow and fast winds. Electric field and heat flux obtained with the model constrained by the observations will be illustrated and discussed in section 5. Concluding remarks will be presented in the last section.

\section{Exospheric model}
The Lorentzian exospheric model has been developed assuming a Kappa distribution for the electrons in ion exospheres \cite{Pierrard-Lemaire-1996P, Maksimovic-etal-1997aP, Pierrard-Pieters-2014P}. The solar wind exospheric model has been recently improved by considering regularized Kappa distributions \cite{Scherer-etal-2017P, Lazar-etal-2020bP} that have no diverging moments through consideration of a cut-off at relativistic velocities. The model becomes valid even for kappa indices lower than 2, which was needed since low values of kappa are sometimes observed in the fast solar wind \cite{VP23}.

Assuming velocity distribution functions of the particles at the exobase, we can determine the distributions at all radial distances, and thus calculate  their different moments \cite{Pierrardbook-2021}. The presence of enhanced suprathermal electron tails accelerates the wind above the exobase in exospheric regions by increasing the electric potential and the flux of escaping electrons. 

\begin{figure}
\includegraphics{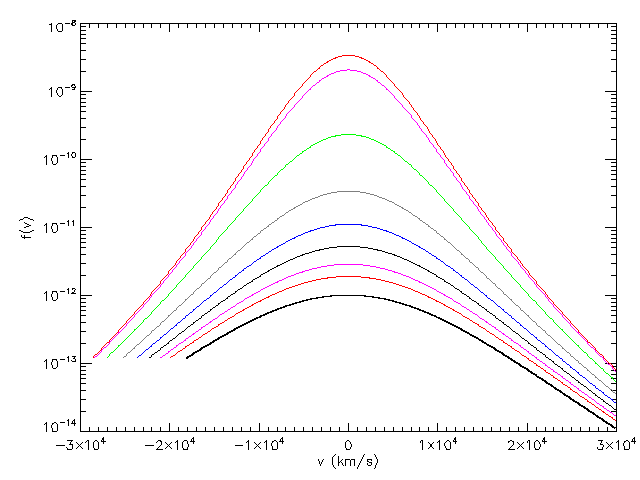}
% Here is how to import EPS art
\caption{\label{fig:epsart2} The electron velocity distribution function in the exospheric model for kappa=4 at different radial distances starting from the exobase at 1.17 Rs (top in red) to 1.2 Rs (pink), 1.5 (green), 2 (grey), 2.5 (blue), 3 (dark grey), 3.5 (purple), 4 (red), and 5 Rs (bottom in black).}
\end{figure}

In the present exospheric model, the distribution is a regularized Kappa function truncated for negative velocities lower than minus the escape velocities \cite{Pierrard-Lemaire-1996P, Maksimovic-etal-2005P}. Indeed, if the electrons escape, they don't come back to the Sun which leads to an empty region of velocity space, i.e., a deficit of particles in the sunward direction. The escape velocity decreases with the radial distance and the suprathermal tails decrease less with the distance than the core, due to the velocity filtration effect \cite{Scudder-1992B, Pierrard-etal-2004P}.  This is well visible on Figure \ref{fig:epsart2} showing the electron truncated Kappa distribution from 1.7 Rs (upper red line) to 5.0 Rs (lower black line), with intermediate distances at 1.2 Rs, 1.5, 2, 2.5, 3, 3.5, and 4.

The most recent observations of PSP confirm such a deficit in the electron distributions at low distances \cite{Bercic2021wh}. They also show the existence of suprathermal electrons even at very low radial distances \cite{Maksimovicbook-2021, Abra23}, which was already observed with Helios  for distances higher than 0.35 AU \cite{Pierrard-etal-2016P, Pierrard-etal-2022}. The quantity of suprathermal electrons is even higher at large distances ($> 1$ AU) as indicated by lower values of the kappa index \cite{Maksimovic-etal-1997bP, Pierrard-Meyer-2017P, Pierrard-etal-2020P}.

\section{Analysis of recent solar wind  observations to constrain the model}
In the present work, we analyze recent data from different spacecraft:

\begin{itemize}
    \item Parker Solar Probe (PSP) from the 24th of January 2020 to the 30th of March 2023 from 13 to 50 R$_s$,
    \item Solar Orbiter (SolO) from the 7th of July 2020 to the 30th of May 2023	from 63 to 218 R$_s$,
    \item OMNI from the 7th of July 2020 to the 30th of May 2023 at 215 R$_s$ (1 AU).
\end{itemize}

The observations from SolO and OMNI are discriminated using wind velocity $u<400$ km/s for the slow wind and $u>500$ km/s for the fast wind.

The criteria chosen for the selection of the fast and slow solar wind case of course influence the averaged values of the moments. The averaged velocity in the fast wind is slightly lower than in \cite{VP23} since $u>$ 600 km/s was chosen for the fast wind selection. 
Moreover, the recent observations of PSP and SolO contain much more fast solar wind cases than in \cite{VP23} since the available data were limited to March 2022. Indeed, the solar activity is in its increasing phase with the next maximum expected in 2025.

For PSP, the criteria have to be different than SolO and OMNI because the observations are located in the acceleration region close to the Sun. This implies that distinguishing fast and slow winds using solely a velocity criterion is not sufficient since a considerable acceleration remains until the wind reaches its final velocity. Halekas et al. (2022) \cite{Halekas-etal-2022} assessed the remaining acceleration by computing the asymptotic speed which requires a measure of the electric potential at each wind velocity measurements. To avoid the need of this electric potential, an indirect way of estimating the remaining acceleration based on the correlation between the wind velocity and the proton temperature is proposed here and will be discussed in the next section. 

\begin{figure}
\includegraphics[width=1\linewidth]{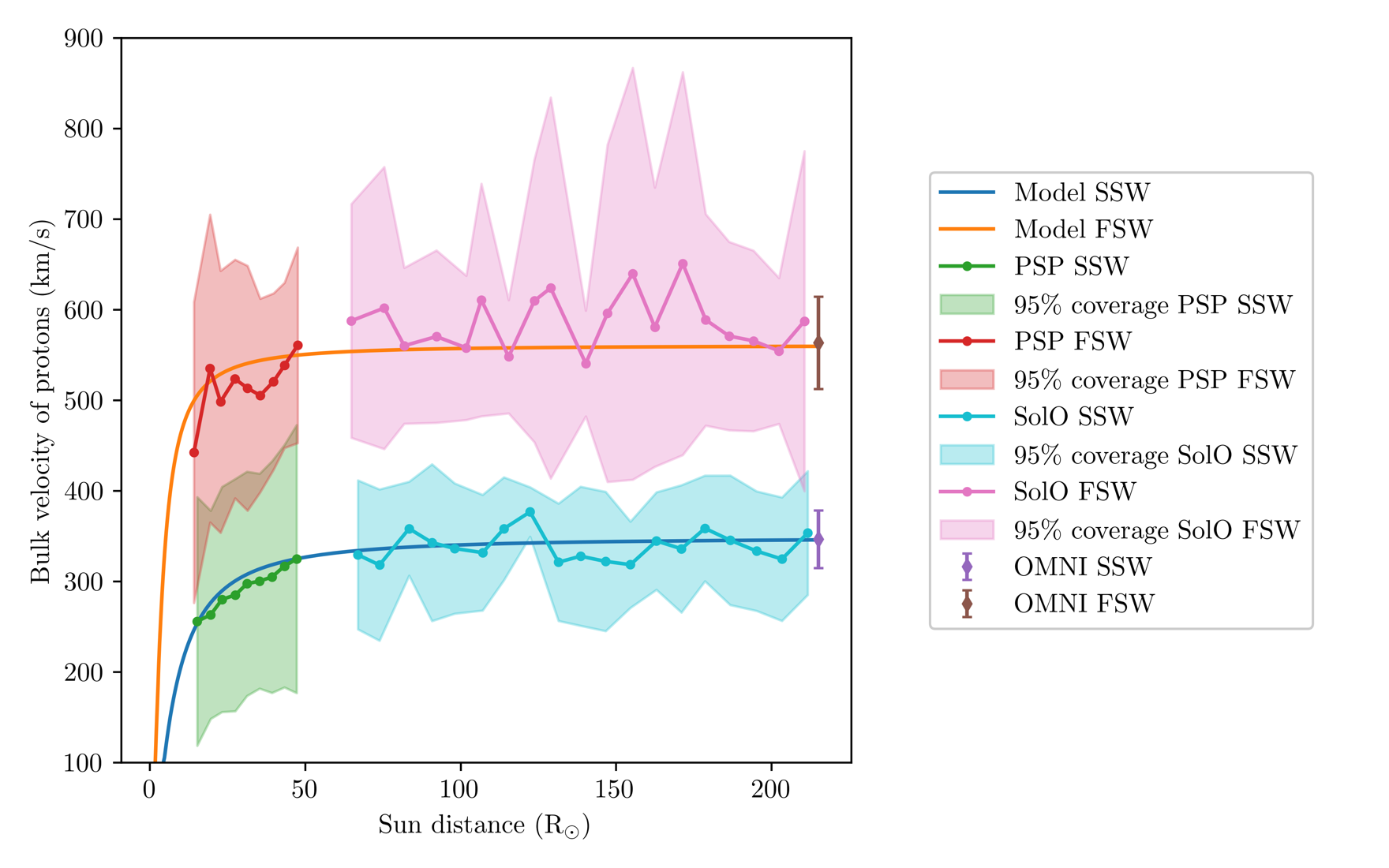}% Here is how to import EPS art
\caption{\label{fig:bulk}Averaged bulk velocity of protons and electrons as a function of the solar distance for the exospheric model constrained by averaged observations of Parker Solar Probe ($<50$ R$_s$), Solar Orbiter ($>63$ R$_s$), and OMNI (215 R$_s$) for fast solar wind (FSW in orange) and slow solar wind (SSW in blue).}
\end{figure}

\begin{figure}
\includegraphics[width=1\linewidth]{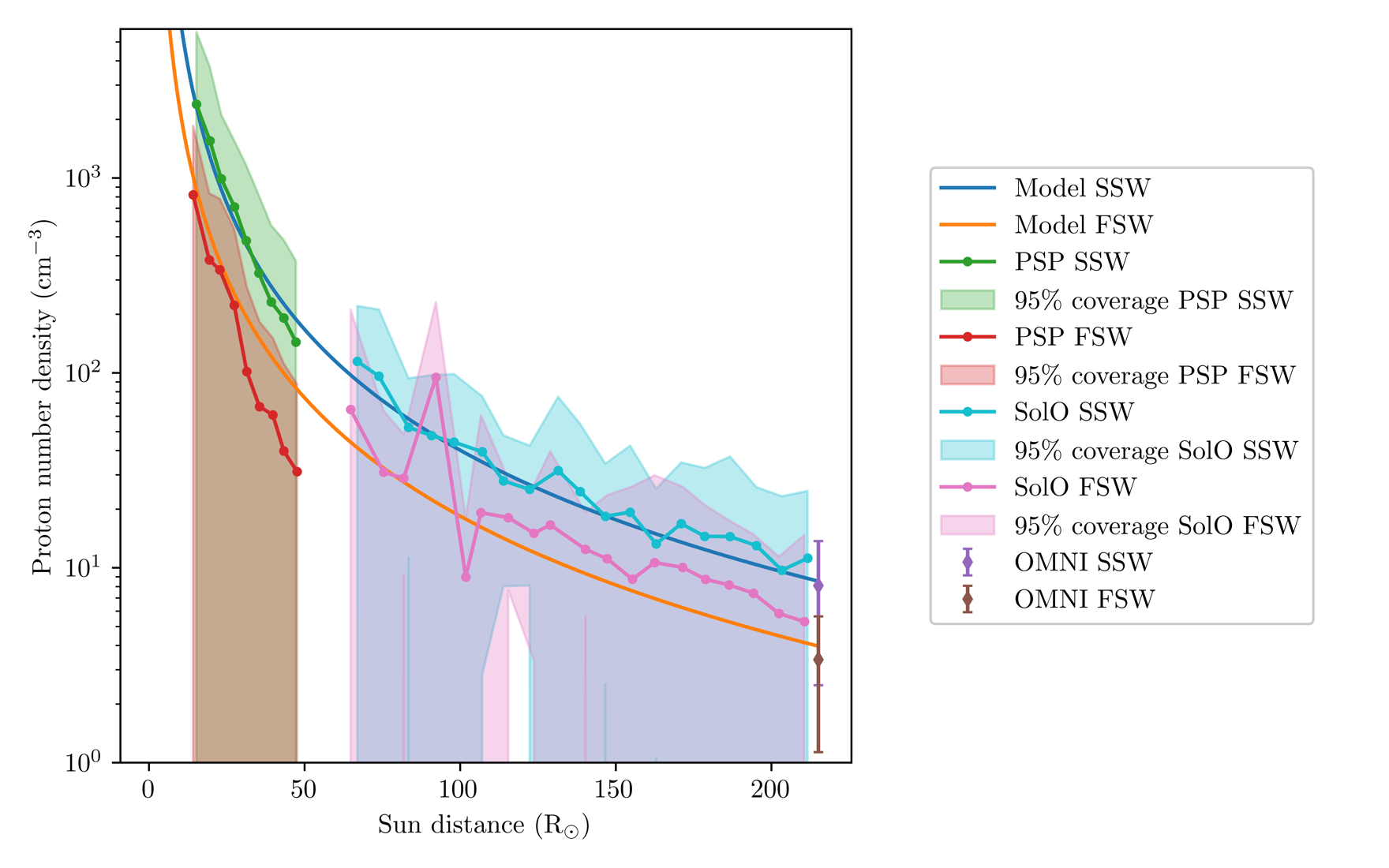}% Here is how to import EPS art
\caption{\label{fig:dens}Averaged number density of protons and electrons as a function of the solar distance for the exospheric model constrained by averaged observations of Parker Solar Probe ($<50$ R$_s$), Solar Orbiter ($>63$ R$_s$), and OMNI (215 R$_s$) for fast solar wind (FSW in orange) and slow solar wind (SSW in blue).}
\end{figure}

\begin{figure}
\includegraphics[width=1\linewidth]{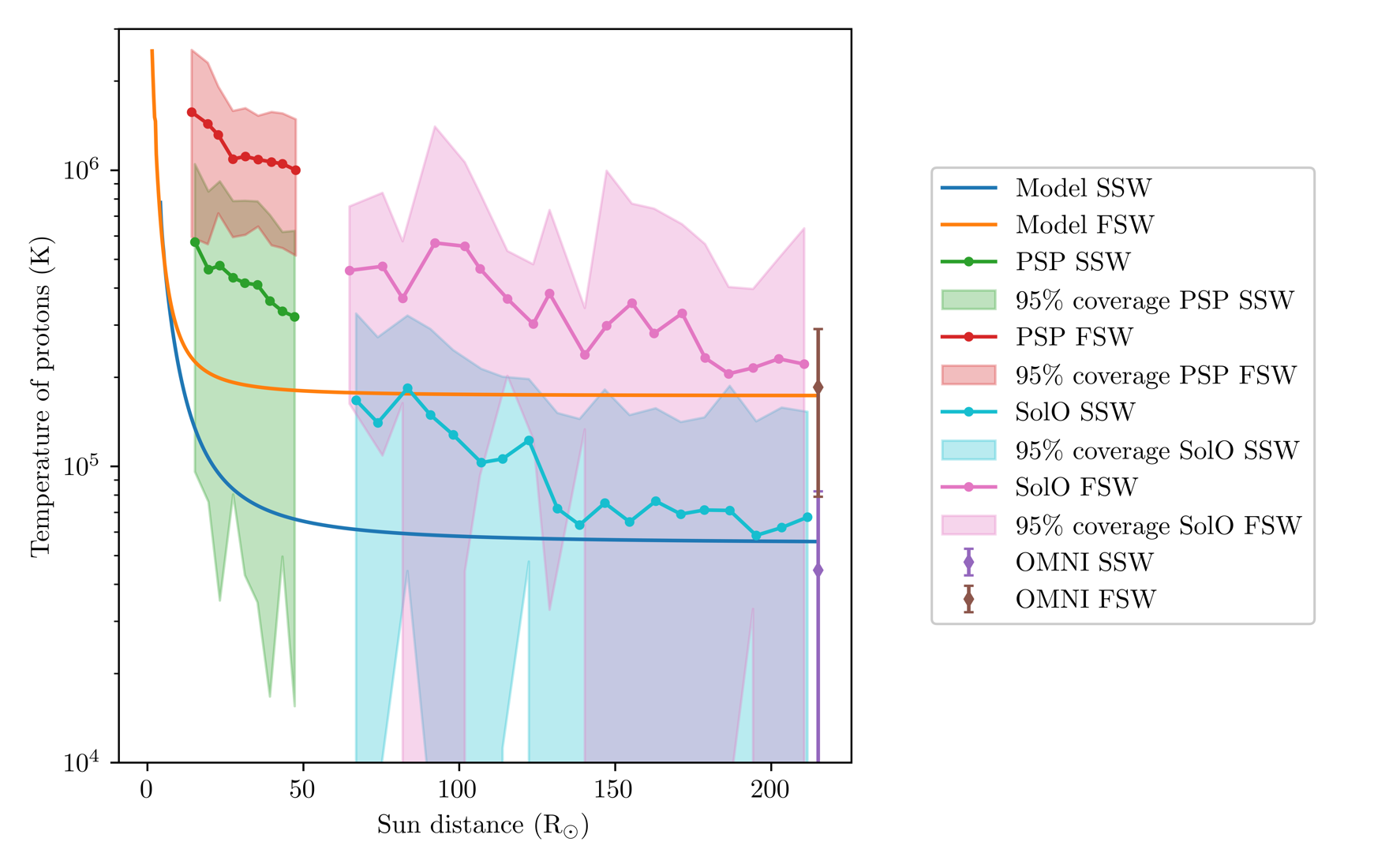}% Here is how to import EPS art
\caption{\label{fig:temp}Averaged proton temperature as a function of the solar distance for the exospheric model constrained by averaged observations of Parker Solar Probe ($<50$ R$_s$), Solar Orbiter ($>63$ R$_s$), and OMNI (215 R$_s$) for fast solar wind (FSW in orange) and slow solar wind (SSW in blue).}
\end{figure}

In Figures \ref{fig:bulk}, \ref{fig:dens}, and \ref{fig:temp}, the points represent the average values of the data (respectively the bulk velocity, density and proton temperature), and the colored area represents the double of the standard deviation interval around average value. This was obtained by binning data for each $4$ R$_s$ interval for PSP  and 8 R$_s$ for SolO and OMNI starting from the lowest registered distance of the corresponding set of data. The average and standard deviation was then computed on each of those bins \cite{Maxp2023}. Note that the gap between the observations of PSP and SolO is due to the lack of sufficiently accurate measurements beyond 50 R$_s$ from PSP.

These figures also show a comparison with the profiles obtained with the exospheric Kappa model where the parameters of the model have been chosen in order to best fit the averaged observed values of Parker Solar Probe (PSP) Solar Orbiter (SolO) and OMNI at 1 AU, respectively in orange for the fast solar wind (FSW) and blue for the slow solar wind (SSW). 
The parameters found for the exospheric model are provided in Table 1.

Figure \ref{fig:bulk} illustrates the model capabilities of producing a realistic bulk velocity by setting the parameters found in Table 1 constrained by PSP, SolO, and OMNI averaged data separated in fast and slow wind.
The main parameters that are responsible for the acceleration of the solar wind are the exobase level and the kappa index. In fact, a lower exobase and/or a lower kappa index will accelerate the wind.
Furthermore, any changes in the exobase's altitude is much more effective when kappa is small. 
It is worth noting that when the fast wind is considered above 500 km/s, the corresponding kappa index is 2.4 (see Table 1), while it was estimated as 2.23 when the fast wind was considered above 600 km/s in \cite{VP23}. 

\begin{table}[b!]
\caption{\label{tab:1}This table gives the parameters used in the exospheric model to best reproduce the Slow (SSW) and Fast Solar Wind (FSW) as observed on average by PSP, SolO, and OMNI up to 1 AU.  }
\begin{ruledtabular}
\begin{tabular}{llllll}
Wind type&Exobase & Exobase density&Electron temperature& Proton temperature& kappa electrons\\
Symbol [units] & $r_0$ [R$_s$] & $n(r_0)$ [m$^{-3}]$ & $T_e(r_0)$ [K]& $T_p(r_0)$ [K]& $\kappa$ \\
\midrule
SSW & 3.9 & $1.2 \times 10^{11}$& $1.5 \times 10^{6}$&$1.25 \times 10^{6}$&  5\\
FSW & 1.25 & $1.4 \times 10^{12}$&$1.35 \times 10^{6}$&$4.06 \times 10^{6}$& 2.4 \\

\end{tabular}
\end{ruledtabular}
\end{table}

Figure \ref{fig:dens} shows the density profiles of the observations and the model. As expected, the density decreases faster in the fast wind than in the slow wind. The density at the exobase is the main parameter that needs to be constrained in the model to best reproduce the observations. Note that the SPAN-I instruments of PSP are not well calibrated to measure low densities \cite{Livi22}, which could explain the apparent discontinuity between PSP and SolO observations that can not be resolved by extrapolating the trend. 

Figure \ref{fig:temp} shows the proton temperature profiles of the observations and the model. The temperature $T_p(r_0)$ at the exobase can be adjusted in the model to best reproduce the observations in the corona. One can see that the temperatures at 1 AU approximately correspond to the observations of OMNI (that are slightly lower than the averaged observations obtained by SolO at this same distance, probably due to the fact that SolO and OMNI does not always observe the same kind of events from their respective locations on the ecliptic plane). Both temperatures decrease too fast in comparison to the observations of SolO and PSP. This is due to the strong assumption of absence of interactions in the exospheric model. It is worth noting that extrapolating the trend of PSP measurements would not resolve the discontinuity with SolO measurements. This is due to the exceptional events that PSP has encountered during its close approaches coupled with the fact that the criteria distinguishing fast and slow wind is not only based on the velocity for PSP data (while it is for SolO data) but is rather a combination between velocity and temperature that will be discussed in the next section.   

\section{The temperature-velocity correlation used as source of identification}
\begin{figure}
\includegraphics[width=.8\linewidth]{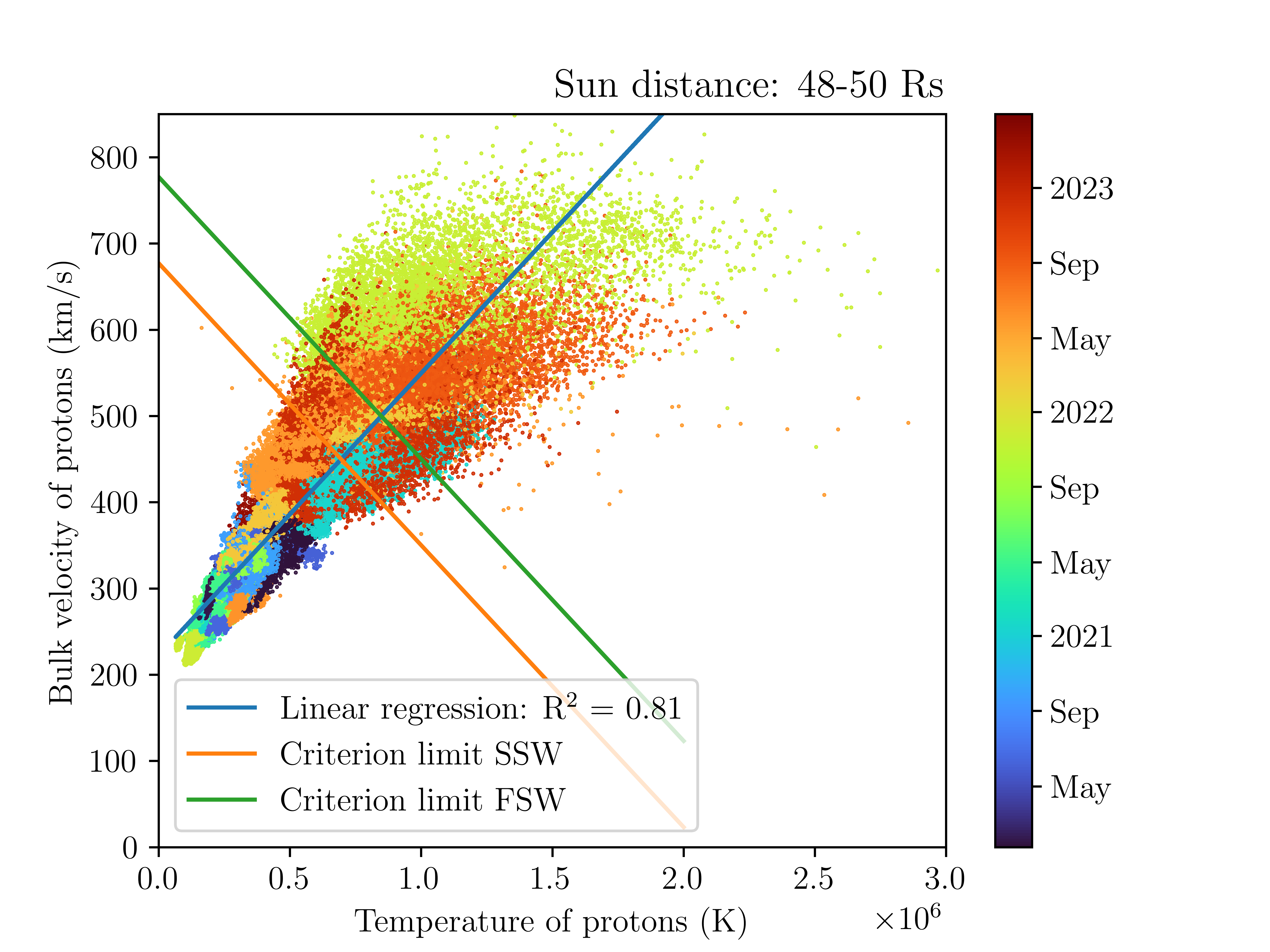}% Here is how to import EPS art
\caption{\label{fig:correl}Proton bulk velocity as a function of the temperature of Parker Solar Probe observations between 48–50 R$_s$ for all orbits from end of January 2020 to end of March 2023, excluding Solar Energetic Particle events.  The linear regression considering the entire displayed data is shown in blue. The perpendicular orange and green lines  respectively delimit the slow and fast winds (see text).}
\end{figure}

As obtained in \cite{Pierrard-etal-2016P} at 1 AU with Helios and CLUSTER and confirmed at lower distances with PSP observations \cite{Maksimovic-etal-2020P}, there is a strong correlation between the proton temperature and the velocity. This correlation is illustrated in Figure \ref{fig:correl} between 48-50 Rs using PSP data. One finds a linear regression (blue line) with a correlation coefficient of 90 $\%$. 

Two Solar Energetic Particle events have been noticed in PSP data, one in early September 2022 and one in mid-March 2023. These events are characterized by higher temperatures than expected, thus deviating from the direct correlation that has been established during multiple other orbits at all distances in PSP data. The data points corresponding to those events have not been considered in Figure \ref{fig:correl} and in aforementioned figures. 

The observed correlation (illustrated by the linear regression in blue line) can be used to identify slow and fast wind at low radial distances where the final velocity is not yet reached. In the case of low radial distance, we define the slow solar wind observed by PSP as the points located below the orange line in Figure \ref{fig:correl}. This line is perpendicular to the linear regression (blue line) and intersects it at  450 km/s. The fast wind correspond to the points above the green perpendicular line crossing the regression at 500 km/s. The points between the two lines are excluded to better discriminate slow and fast winds. This gives better results than just separating the velocities below 450 km/s and above 500  km/s at low distances. It is here assumed that most of the winds with sufficiently high proton temperature (and velocity) will have a significant remaining acceleration, allowing them to be considered as fast solar winds where they would sometimes have been considered slow if a simple velocity criterion was considered.   

\section{Electric field in hydrodynamic and exospheric models}

Parker [1958] \cite{Park58} was the first one to propose an hydrodynamic model for the solar atmosphere, giving up the hypothesis of hydrostatic equilibrium assumed previously. The energy is evacuated by the radial expansion of the corona, leading to supersonic speeds of the solar wind. Magnetohydrodynamic (MHD) models assume collision-dominated solar wind and are also able to reproduce many solar wind characteristics. Nevertheless, the hypothesis of plasma dominated by collisions is not valid above the exobase. This highlights the complementarity of the MHD approach at low distances and the kinetic exospheric approach at larger distances, which can be combined for optimal solar wind results \cite{Moschou-etal-2017P}.

It is important to note that the effect of the electric field $E$ is included in MHD hydrodynamic models as well \cite{Pierrard-2009aP}. Indeed, in the famous Parker model for instance \cite{Park58}, the hydrodynamic equations for electrons (index $e$) and protons (index $p$) are: 
\begin{eqnarray}
&&n_e m_e u_e\frac{\partial u_e}{\partial r}+ \frac{\partial p_e}{\partial r}=-n_e m_e g -n_e e E \label{nel}\\
&&n_p m_p u_p\frac{\partial u_p}{\partial r}+ \frac{\partial p_p}{\partial r}=-n_p m_p g +n_p e E
\label{npro}
\end{eqnarray}
where $p$ is the pressure, $m$ the mass of the particle, $g$ the gravitational acceleration, and $e$ the electric charge.

 By considering the following hypotheses:   
\begin{itemize}

    \item Quasi-neutrality: $n_e=n_p$

\item No electric current: $u_p=u_e$
\end{itemize}

\noindent the global equation is then written by adding equations (\ref{nel}) and (\ref{npro}):

\begin{eqnarray}
\rho u\frac{\partial u}{\partial r}+\frac{\partial p}{\partial r}=-\rho g
\label{con}
\end{eqnarray}
where $\rho$ is the mass density.

The electric field has thus disappeared from the global equation of the solar wind plasma, but this does not mean that the electric field is null. More specifically, it is here ensuring the equality of the proton and electron fluxes so that there is no net current, just like in kinetic models. 

One of the advantages of the kinetic exospheric approach is its ability to clearly emphasize that the acceleration of the solar wind is due to the electric force, reducing the gravitational attraction of the protons and heavy ions \cite{Pierrard-Lamy-2003P}. It is important to highlight that in the kinetic exospheric approach, the electric field is much higher than the Pannekoek-Rosseland \cite{PR22} electric field, because of the presence of escaping particles \cite{Lemaire-Pierrard-2001P}. Moreover, it allows to take into account suprathermal electrons using Kappa distributions \cite{Pierrard-2012aP}.

\begin{figure}
\includegraphics[width=.8\linewidth]{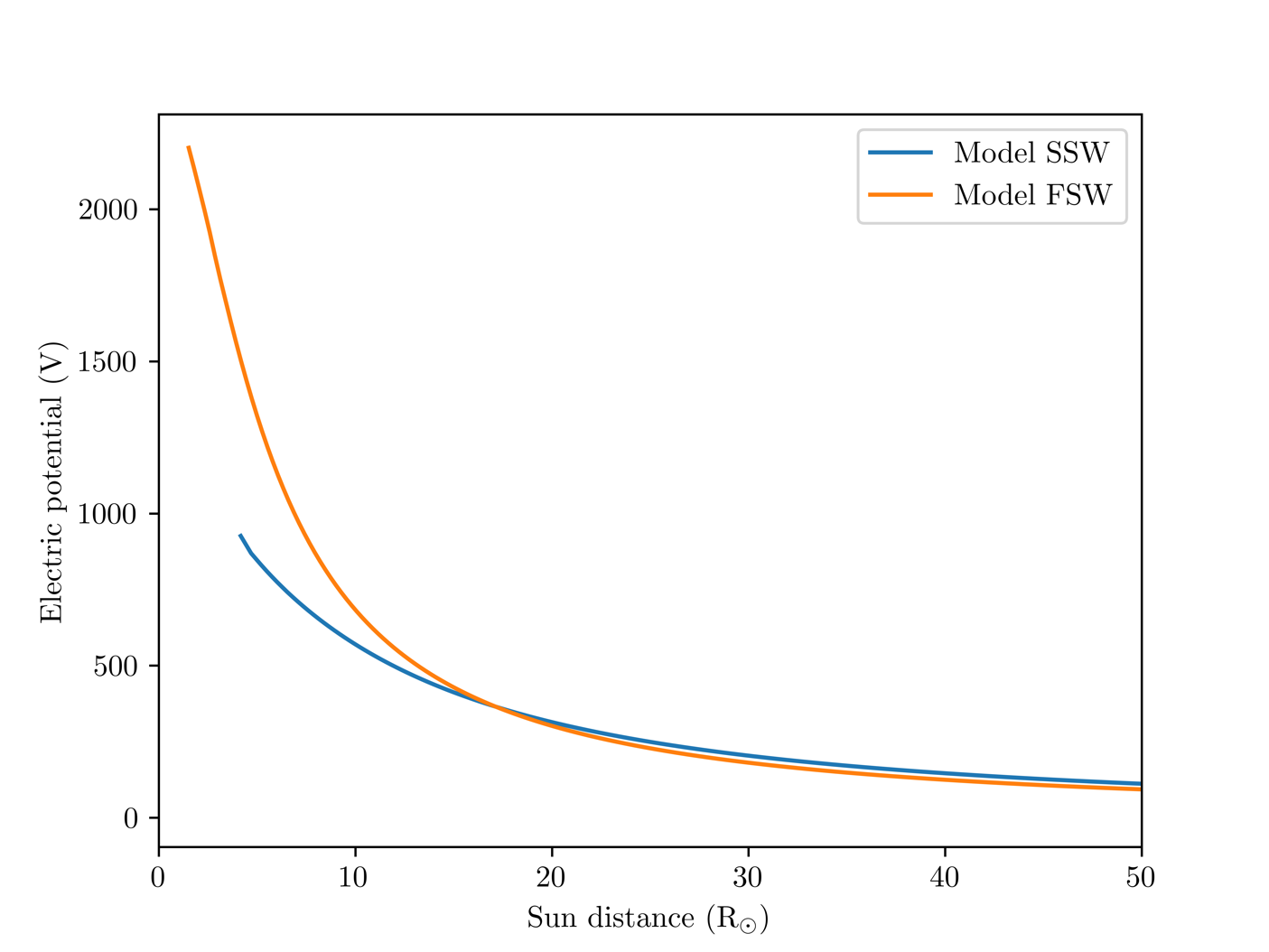}% Here is how to import EPS art
\caption{\label{fig:elec}Electric potential  obtained with the exospheric model using the parameters of Table 1 for the fast (orange) and slow wind (blue).}
\end{figure}

Figure \ref{fig:elec} illustrates the electric potential profiles found for the SSW and FSW with the Kappa exospheric model with the parameters of Table 1. To be accelerated to higher bulk velocities, the fast wind needs a higher potential difference between the exobase and the large radial distances. This is obtained by considering a lower exobase or by decreasing the kappa index. Halekas et al. (2022)  \cite{Halekas-etal-2022} measured that the  electric potential is higher in the slow wind than in the fast wind above 14 R$_s$. This is indeed obtained also in the model, as illustrated in Figure \ref{fig:elec}, due to the low exobase for the fast wind and a sufficiently higher exobase for the slow wind. The difference is rather slight in Figure \ref{fig:elec} but can be increased considering a higher exobase for the slow wind, which does not considerably modify the SSW profile. 

\begin{figure}
\includegraphics[width=.8\linewidth]{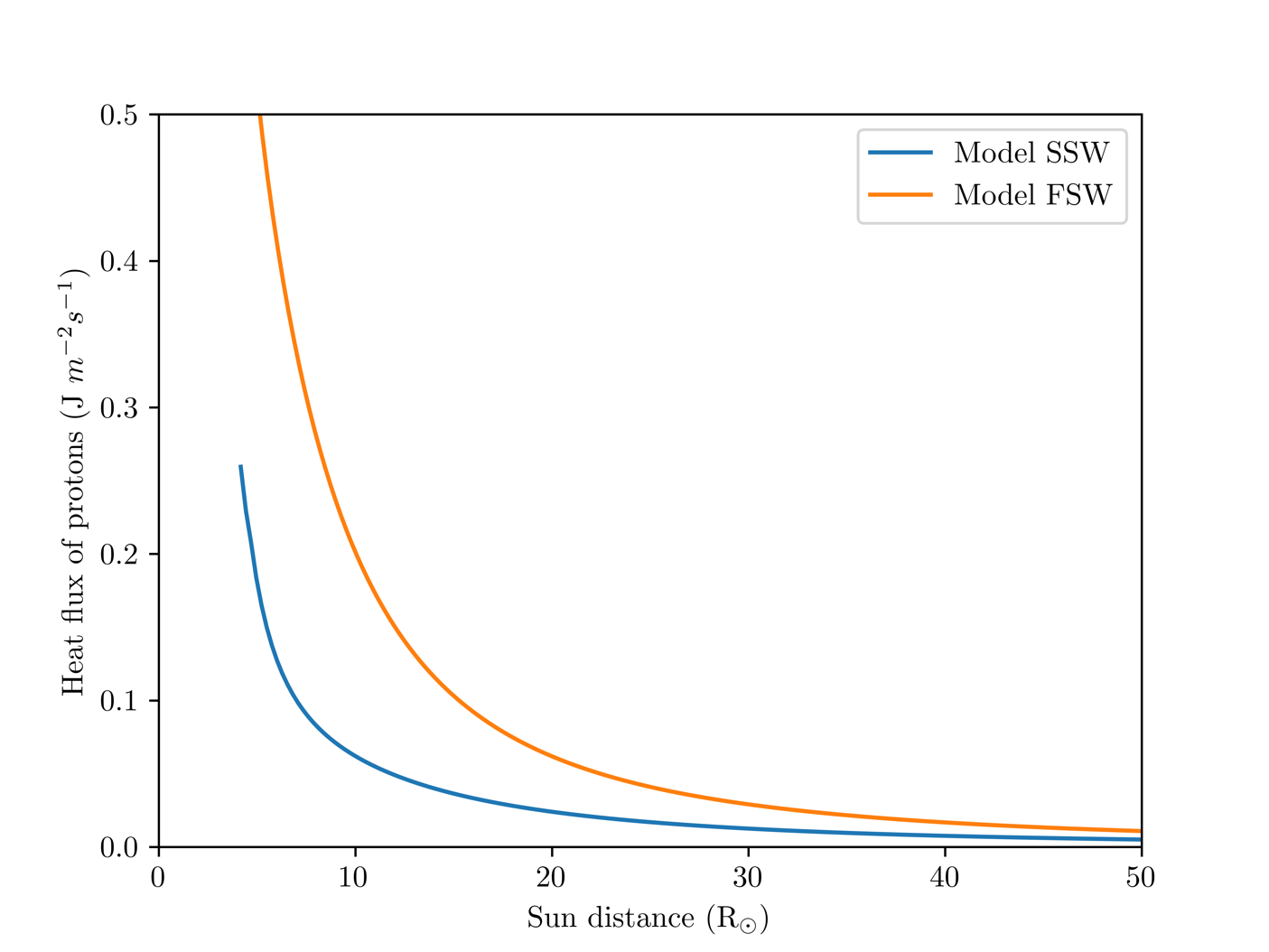}% Here is how to import EPS art
\caption{\label{fig:hflux}Heat flux  obtained with the exospheric model using the parameters of Table 1 for the fast (orange) and slow wind (blue).}
\end{figure}

As illustrated in Figure \ref{fig:hflux}, the heat flux obtained with the model  decreases with the radial distance, as observed with PSP \cite{Halekas-etal-2020P}.   It is  higher in the fast wind than in the slow wind. The heat flux produced by the model is  too high in comparison to observations \cite{Pierrard-2012bP, VP23}. This is expected since exospheric models give a  maximum level of the possible heat flux. In fact, neither purely collisionless models nor purely collisional mechanisms are able to explain the  heat flux decay as measured in the solar wind \cite{Bale13}. Whistler waves, commonly observed in the interplanetary space, are a good candidate to solve the problem of heat flux regulation \cite{Colo23}. Effects of whistler wave turbulence on the evolution of the electron distribution function in the solar wind were shown to be able to modify the suprathermal populations \cite{Pierrard-etal-2011P}, and especially spread the strahl to the halo as observed in the solar wind \cite{Stve9}. Other plasma instabilities can also help to understand the characteristics of the observed distributions, and especially their temperature anisotropies and heat fluxes \cite{Sun21, Vers22}.

For proton distributions, kinetic Alfvén wave turbulence presents another potential mechanism in the still-debated energy responsible for accelerating plasma and for the formation of the proton beam \cite{Pierrard-Voitenko-2013P, Voit15}

\section{Conclusion}

In this section, we summarize the main results of the present study.

\begin{itemize}
    \item More precise boundary conditions were found for the exospheric model by constraining it with PSP, SolO and OMNI observations from their launch to mid-2023. The values of the radial distance of the exobase, density, temperature and kappa index at the exobase for the slow and fast wind were determined to best fit the averaged profiles of the moments.
    \item The criterion distinguishing fast and slow wind was improved in the case of a significant remaining acceleration. This improvement shows that the acceleration of the solar wind seems to closely follow what was predicted by the exospheric model. This was achieved only by deducing the criteria from the observed correlation between wind speed and proton temperature.
    \item The exobase altitude also has important implications in the acceleration of the solar wind. This implies that the electric field can still be higher for slower winds if the exobase is substantially higher than for faster winds.
    \item It has been shown that the electric field is important also in MHD models. 
    \item One of the main advantages of kinetic models is their ability to consider suprathermal particles.
    \item The exospheric models give a maximum heat flux, which can be improved by considering interactions between the particles. Fokker-Planck models including Coulomb collisions and wave-particle turbulence provide significant improvement concerning the regulation of the heat flux  and the temperature anisotropy.   
\end{itemize}

\begin{acknowledgments}
The project 21GRD02 BIOSPHERE has received funding from the European Partnership on Metrology, co-financed by the European Union’s Horizon Europe Research and Innovation Programme and by the Participating States.
\end{acknowledgments}

\nocite{*}
\bibliography{aipsamp}% Produces the bibliography via BibTeX.

\end{document}